\documentclass[runningheads]{llncs}
\usepackage{graphicx}

\usepackage{tikz}
\usepackage{comment}
\usepackage{amsmath,amssymb} 
\usepackage{color}
\usepackage{hyperref}
\hypersetup{colorlinks=true, urlcolor=magenta}

\begin{document}
\pagestyle{headings}
\mainmatter
\def\ECCVSubNumber{106}  

\title{PyNET-CA: Enhanced PyNET with Channel Attention for End-to-end Mobile Image Signal Processing} 

\titlerunning{PyNET-CA: Enhanced PyNET with Channel Attention}
%
\author{Byung-Hoon Kim\inst{1} \and
Joonyoung Song\inst{1} \and
Jong Chul Ye\inst{1} \and
JaeHyun Baek\inst{2}}
%
\authorrunning{B. -H. Kim et al.}
%
\institute{Korea Advanced Institute of Science and Technology, Daejeon, South Korea \\
\email{\{egyptdj,songjy18,jong.ye\}@kaist.ac.kr} \and
Amazon Web Services, Seoul, South Korea\\
\email{jakemraz100@gmail.com}}
\maketitle

\makeatletter
\def\blfootnote{\xdef\@thefnmark{}\@footnotetext}
\makeatother

\begin{abstract}
Reconstructing RGB image from RAW data obtained with a mobile device is related to a number of image signal processing (ISP) tasks, such as demosaicing, denoising, etc.
Deep neural networks have shown promising results over hand-crafted ISP algorithms on solving these tasks separately, or even replacing the whole reconstruction process with one model.
Here, we propose PyNET-CA, an end-to-end mobile ISP deep learning algorithm for RAW to RGB reconstruction.
The model enhances PyNET, a recently proposed state-of-the-art model for mobile ISP, and improve its performance with channel attention and subpixel reconstruction module.
We demonstrate the performance of the proposed method with comparative experiments and results from the AIM 2020 learned smartphone ISP challenge.
The source code of our implementation is available at \url{https://github.com/egyptdj/skyb-aim2020-public} \blfootnote{The final authenticated version is available online at \url{https://doi.org/10.1007/978-3-030-67070-2_12}}

\end{abstract}

\begin{figure}
\centering
\includegraphics[width=1.0\textwidth]{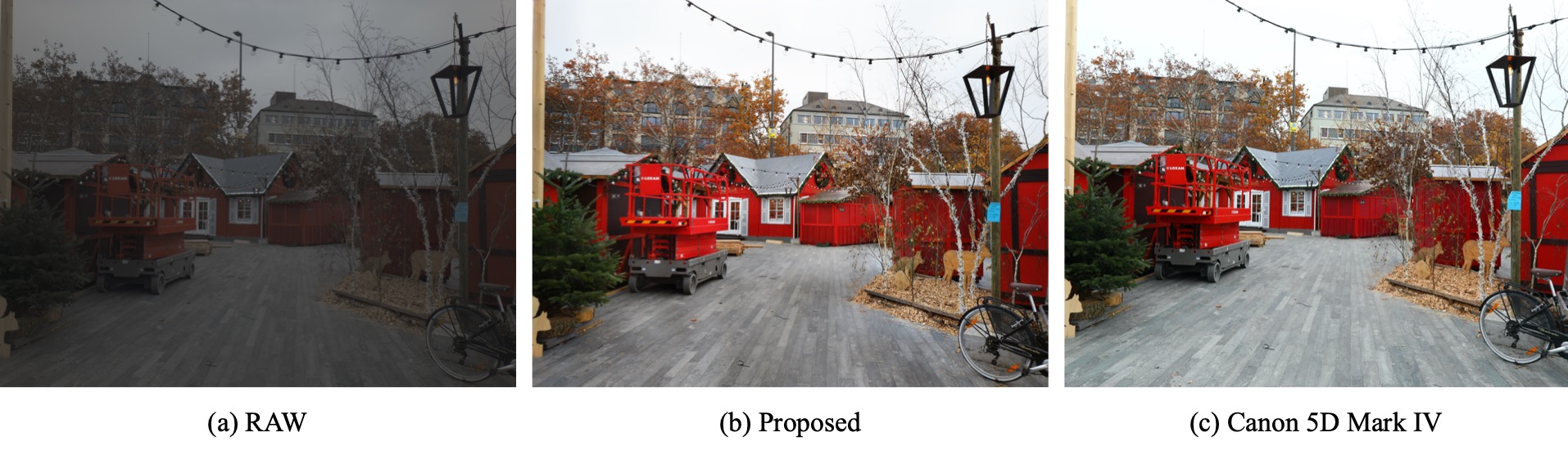}
\caption{Reconstructed RGB image from RAW data with the proposed method. (a) Input RAW image (visualised). (b) Reconstructed RGB image with proposed method. (c) Target image taken with Canon 5D Mark IV.}
\label{fig:rep}
\end{figure}

\section{Introduction}
Reconstructing RGB image from the RAW data obtained with a mobile device is a topic of growing interest.
The data acquired from the image sensors of mobile devices require several image signal processing (ISP) steps to solve a number of low-level computer vision problems, such as demosaicing, denoising, color correction, etc.
Hand-crafted ISP algorithms depend on the prior knowledge about the data acquisition process or degradation principles.
Softwares of the mobile device implement these algorithms to process the RAW data sequentially, solving each tasks step by step to reconstruct the RGB image presented to the user.

Deep neural networks, specifically the convolutional neural networks (CNNs), have recently shown promising results over hand-crafted ISP algorithms.
There also have been attempts to not just replace each ISP algorithm with deep neural networks, but to reconstruct RGB images from the RAW data by training a single end-to-end deep learning reconstruction model.
However, one of the difficulties in training an end-to-end reconstruction model over separately processing the RAW data is that the prior knowledge about the data is not directly incorporated into the model.
For example, it is important to take both global (e.g. luminance, color balance) and local features (e.g. fine-grained textures, edge structures) of the image into account during the reconstruction process, which is not explicitly present in the input RAW data.

We consider a recently proposed end-to-end RAW to RGB reconstruction model PyNET \cite{ignatov2020replacing}, which is basically a CNN designed to exploit both global and local features of the input data.
Despite the fact that the PyNET achieves state-of-the-art performance in RAW to RGB reconstruction, there exists some drawbacks in the model architecture and the training process that can be further improved.
In this paper, we address these issues and propose PyNET-CA, an enhanced PyNET with channel attention to improve the performance and reduce the training time.
We demonstrate the performance of the proposed model by a number of comparative experiments, and report the results of participating the AIM 2020 learned smartphone ISP challenge.

\section{Related Work}
Since the introduction of the SRCNN \cite{dong2015image} which solves the single image super-resolution (SISR) problem with a CNN, a large variety of neural network based image reconstruction and enhancement methods have been proposed.
Deep learning for SISR has rapidly grown with deeper network architectures \cite{kim2016accurate}, and better modules \cite{lim2017enhanced}, \cite{zhang2018residual}.
Models that employ the channel attention mechanism have also shown to improve the performance of image enhancement tasks \cite{zhang2018image}.
Although these development of network architectures do improve the quantiative quality of the enhanced images, it does not necessarily mean that the enhanced images are perceptually of good quality.
Generative adversarial network (GAN) based SISR models, such as SRGAN \cite{ledig2017photo} and ESRGAN \cite{wang2018esrgan}, are introduced to address this issue, and enhances the images in a photo-realistic way.
Enhancing the image to both quantitatively accurate and perceptually realistic within the perception-distortion tradeoff \cite{blau2018perception} is now an important issue in image enhancement tasks \cite{deng2019wavelet}.

Along with the development of neural networks for image enhancement tasks, there have also been attempts to train an end-to-end deep learning model for RAW to RGB reconstruction.
One of the earliest model was proposed by \cite{ignatov2017dslr} which is based on a CNN with composite loss function and adversarial training scheme.
The DeepISP \cite{schwartz2018deepisp} also is based on the CNN structure, and addresses the issue of global and local feature priors with a two-stage approach.
The W-Net is another model with the two-stage approach proposed by \cite{uhm2019w}, which stacks two U-Net \cite{ronneberger2015u} structure with channel attention module.
The SalGAN \cite{zhao2019saliency} employs the U-Net structure as the generator of the adversarial training scheme and incorporates spatial attention scheme into the loss function.
The HERN \cite{mei2019higher} modifies the channel attention module of the residual in residual module of \cite{zhang2018image} to construct a dual-path network and incorporates global feature information with a separate full-image encoder.
One of the most recent model is PyNET \cite{ignatov2020replacing}, a CNN with inverted pyramidal structure.
The PyNET achieves state-of-the-art result on the RAW to RGB reconstruction task, owing to the model architecture that can account for both global and local features of the image \cite{ignatov2020replacing}.

\section{Proposed Method}

\subsection{Network architecture}

\begin{figure}
\centering
\includegraphics[width=10cm]{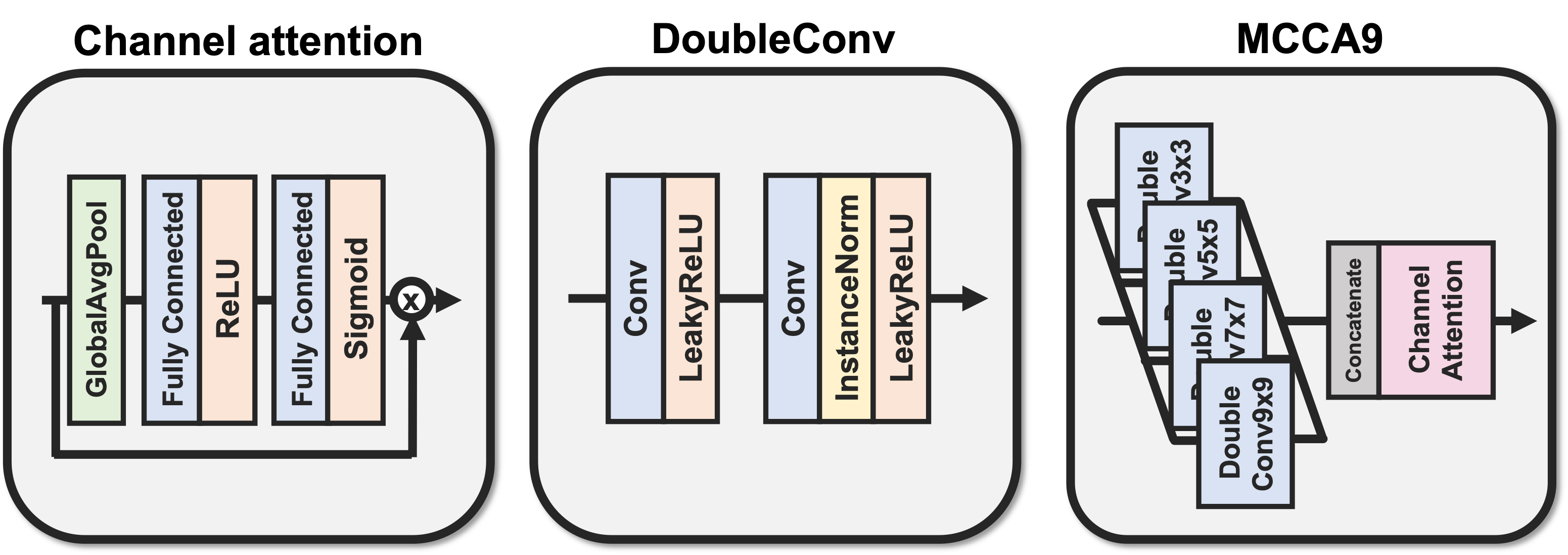}
\caption{Basic modules of the proposed model}
\label{fig:modules}
\end{figure}

\subsubsection{Basic modules}
First, we define some of the modules that constitute the PyNET-CA model (Figure \ref{fig:modules}).
The channel attention module (CA) of PyNET-CA follows from \cite{zhang2018image}. Global average pooling is first applied to the height and width dimension of the features to output a vector with length corresponding to the number of input channels. The pooled vector is linearly mapped, passed through the nonlinear ReLU activation, and is again linearly mapped to match the number of input channels. The length of output features after the first linear mapping is determined by the reduction ratio $r$, which reduces the number of channels by $\frac{1}{r}$ of the input channels. The final features are squashed to range $[0,1]$ by the sigmoid function, and are multiplied element-wise with the channels of the input to account for the level of attention for each channels.

The DoubleConv module is defined as two sequential operations of 2D convolution followed by a LeakyReLU activation.
The kernels of the two convolution layers have the same shape, which is one of $3 \times 3$, $5 \times 5$, $7 \times 7$, $9 \times 9$.
The input image or feature to the DoubleConv module is reflect-padded before the convolution to match the size of the output feature.
Unlike the original PyNET \cite{ignatov2020replacing}, we apply instance normalisation after only the second convolution layer where needed.

The MultiConv channel attention (MCCA) module is the basic building block of our model.
It comprise concatenating the features from the DoubleConv modules and a channel attention module.
We specify four types of MCCA (MCCA3, MCCA5, MCCA7, MCCA9) based on the kernel size of the DoubleConv module.
The MCCA3 module has only one $3\times 3$ DoubleConv module, while the MCCA5 module concatenates the output of $3\times 3$ and $5\times 5$ DoubleConv module, and so forth.
Lastly, channel attention is applied to the concatenated features from the DoubleConv modules.
The reduction ratio $r$ of the channel attention module is set to 1, 2, 3, 4 for MCCA3, MCCA5, MCCA7, MCCA9, respectively.

\begin{figure}
  \centering
  \includegraphics[height=8.2cm]{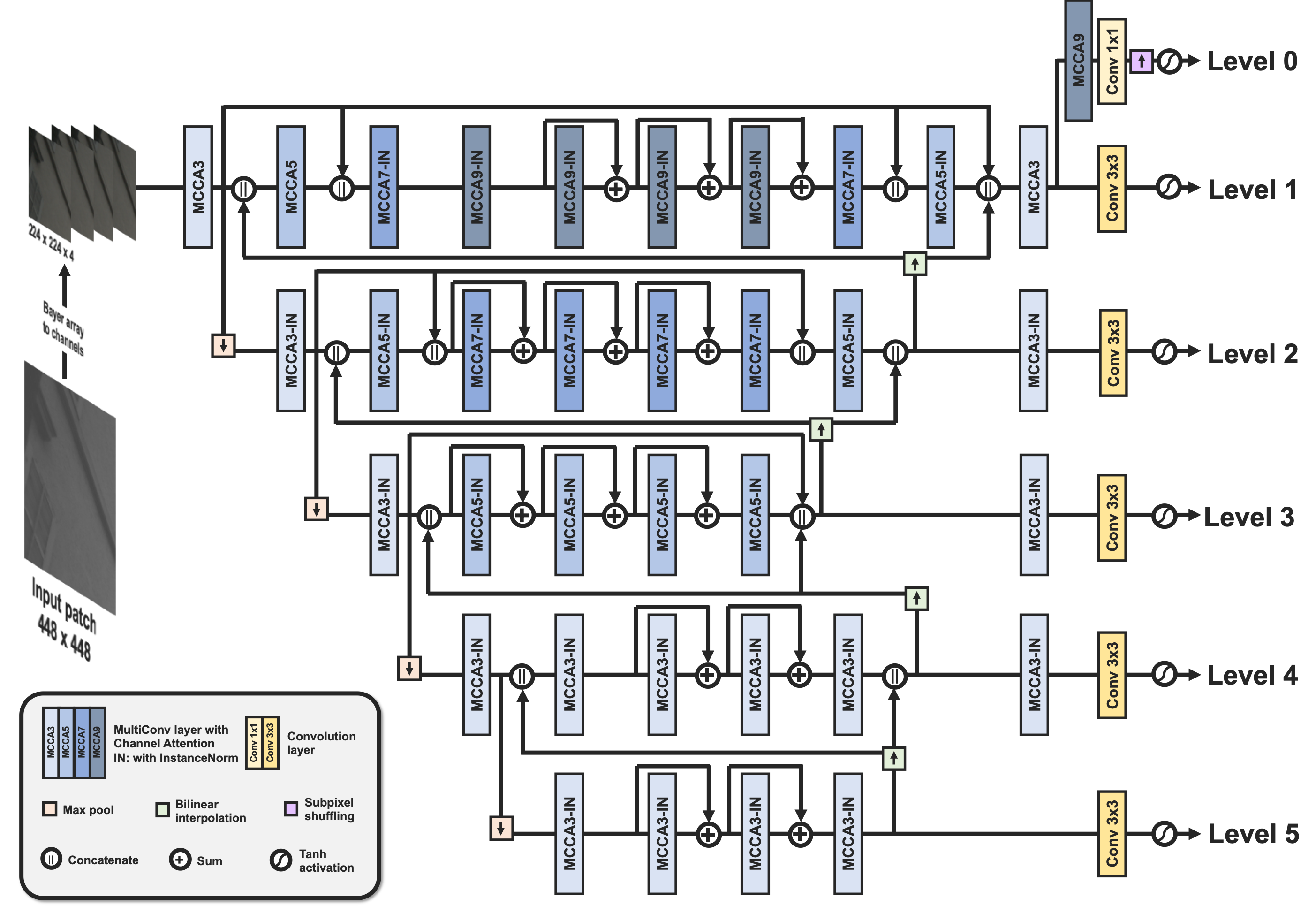}
  \caption{Schematic illustration of the PyNET-CA model.}
  \label{fig:model}
\end{figure}

\subsubsection{Inverted pyramidal structure}

To account for both the global and local features of the image, PyNET-CA has an inverted pyramidal structure as in Figure \ref{fig:model}.
Given a RAW image with Bayer pattern $\mathbf{I}_{raw} \in \mathbb{R}^{H \times W}$, Bayer sampling function $f_{Bayer}$ is first applied to obtain
\begin{align}
  \label{eq:bayer}
  \hat{\mathbf{I}}_{raw} = f_{Bayer}(\mathbf{I}_{raw}) \in \mathbb{R}^{\frac{H}{2} \times \frac{W}{2} \times 4}.
\end{align}

At each levels, the images are downscaled by $2 \times 2$ max pooling followed by the MCCA3 module to serve as the input feature $F_{in}^{k}$ at level $k$,
\begin{align}
  F_{in}^{1} &= \text{MCCA3}(\mathbf{\hat{I}}_{raw}), \\
  F_{in}^{k+1} &= \text{MCCA3}(\text{MaxPool2}(F_{in}^{k})), \quad k\in \{1,2,3,4\}.
\end{align}

We denote the set of operations at each level $k$ that process the input feature $F_{in}^{k}$ as function $H^{k}$.
The function $H^{k}$ is composed of of a number of MCCA modules along with residual connections, within-level and between-level skip connections to compute the output feature $F_{out}^{k}$,
\begin{align}
  F_{out}^{5} & = H^{5}(F_{in}^{5}) \\
  \label{eq:prior}
  F_{out}^{k} & = H^{k}(F_{in}^{k}, F_{out}^{k+1}), \quad k\in \{1,2,3,4\}.
\end{align}
The composition of the operators of $H^{k}$ differs at each levels (Figure \ref{fig:model}).

Lastly, the reconstructed RGB image $\mathbf{\hat{I}}_{rgb}^{k}$ is obtained by a $3 \times 3$ convolution layer and tanh activation of the output features,
\begin{align}
  \mathbf{\hat{I}}_{rgb}^{k} = \text{tanh}(\text{Conv3}(F_{out}^{k})) \in \mathbb{R}^{\frac{H}{2^{k}} \times \frac{W}{2^{k}} \times 3}, \quad k\in \{1,2,3,4,5\}.
\end{align}

One important point regarding the inverted pyramidal structure lies in \eqref{eq:prior}.
It should be noted that the output feature at level $k$ is computed not only with the input feature at level $k$, but also with the output feature from lower-resolution level $k+1$ as a prior.
Because the network is trained progressively, the prior $F_{out}^{k+1}$ contains meaningful information of the downsampled target image.
The bilinear downscaling of the target image at each level corresponds to low-pass filtering, and explicitly emphasizes global feature information as the levels elevate.
This information can be passed onto the lower levels effectively and recursively by concatenation.

\subsubsection{Subpixel reconstruction module}
Because of the Bayer sampling function in \eqref{eq:bayer}, the enhanced features need to be upsampled at the last level of the PyNET-CA.
This is an ill-posed problem as in the case of the SISR problems.
At the final level of the original PyNET structure, enhanced features are upsampled with bilinear interpolation or transposed convolution, and then convolved with a $3 \times 3$ kernel convolution layer to output the final image.
However, bilinear interpolation or transposed convolution can cause blurring or checkerboard artifacts in the upsampled image.
Furthermore, reconstructing final image with the convolution layer after upsampling the image is computationally inefficient.

For computational efficiency and better image quality, the proposed PyNET-CA upsamples the image with the MCCA9 module, followed by $1 \times 1$ convolution layer and upsamples the features by subpixel shuffling \cite{shi2016real} at the final level of the model,
\begin{align}
  \label{eq:srm}
  \mathbf{\hat{I}}_{rgb} = \text{tanh}(\text{SubpixelShuffle}(\text{Conv1}(\text{MCCA9}(F_{out}^{1})))) \in \mathbb{R}^{H \times W \times 3}.
\end{align}
We denote the RGB reconstruction module \eqref{eq:srm} as the subpixel reconstruction module (SRM).

\subsection{Network training}
\begin{figure}
\centering
\includegraphics[width=10cm]{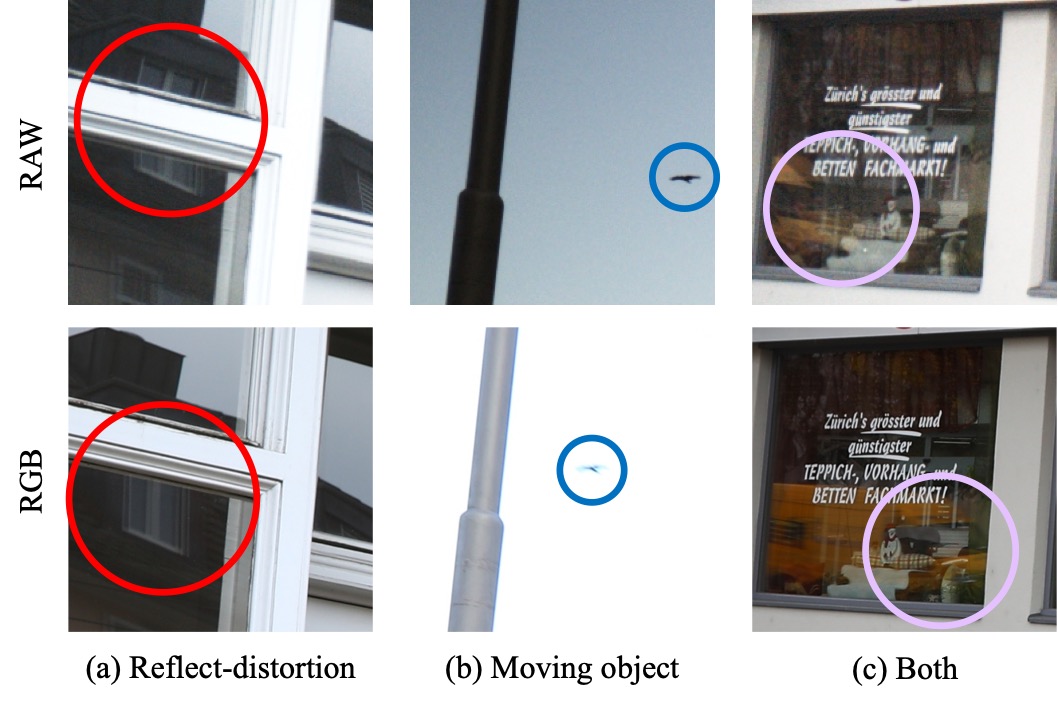}
\caption{Exemplar images from the ZRR dataset with pixel-level mismatch between the RAW and RGB image pairs. The mismatch is due to (a) angular distortion of the reflection, (b) rapidly moving object, (c) reflect-distortion of the rapidly moving object (both).}
\label{fig:dataset_mismatch}
\end{figure}

\subsubsection{Dataset}
For training the network, we used the Zurich RAW to RGB (ZRR) dataset.
One important fact about the dataset is that it consists of a RANSAC \cite{fischler1981random} aligned pair of RAW and RGB images, taken separately from a smartphone (Huawei P20) and a DSLR camera (Canon 5D Mark IV), respectively.
Considering the collection process, the dataset is prone to significant pixel-level mismatch between the RAW and RGB image pairs if the object is moving or if the object has large area of reflection (see Figure \ref{fig:dataset_mismatch}).
We screened out the images with large area of reflection (e.g. on cars, on windows) or moving objects (e.g. people, animal, vehicles on the road).
There were 1,679 out of 46,839 (3.58\%) training image pairs excluded from this screening, leaving out 45,160 image pairs for training the model.
For testing, we used 1,204 image pairs provided by the AIM 2020 challenge organizers without excluding any image pairs.
All RAW and RGB images were $448 \times 448$ cropped patches, and were casted into single precision floating point data centered and scaled to the range $[-1, 1]$ to stabilise the training.

\subsubsection{Progressive training}

The PyNET-CA model is progressively trained from the level with the lowest resolution.
The target image $\mathbf{I}_{rgb}$ is downsampled with bilinear interpolation to match the resolution of the reconstructed images $\mathbf{\hat{I}}_{rgb}^{k}$ at each level.
The loss function used for training the PyNET-CA is a linear combination of the mean squared error (MSE) loss, the perceptual loss using one VGG layer \texttt{relu5\_4}, and the multi-scale structural similarity index measure (MS-SSIM) loss,
$$
  \mathcal{L} = \lambda_{1} \mathcal{L}_{\texttt{MSE}} +\lambda_{2} \mathcal{L}_{\texttt{VGG}} + \lambda_{3} \mathcal{L}_{\texttt{MS-SSIM}}.
$$

Basically the MSE loss is minimised at all levels with $\lambda_{1}$ set to 1.0, and other coefficients were adjusted with respect to the $\lambda_{1}$.
For training level 5 and level 4, only the MSE loss is minimised, with $\lambda_{2}$ and $\lambda_{3}$ set to 0.0.
From level 3, the perceptual loss is minimised to account for the perceptual similarity with coefficient $\lambda_{2}=0.01$, $\lambda_{3}=0.0$.
At the last 0th level, the MS-SSIM loss is maximised (i.e., negatively minimised).
Different MS-SSIM scaling coefficient is used for training the model based on the goal of the task.
For the fidelity task, which is to achieve highest metric score on the peak signal-to-noise ratio (PSNR) and the MS-SSIM, the MS-SSIM scaling coefficient $\lambda_{3}$ is set to 0.01.
For the perceptual task, on the other hand, we set the MS-SSIM scaling coefficient to 0.1.

\subsubsection{One-cycle policy of the learning rate}
\begin{figure}
\centering
\includegraphics[width=10cm]{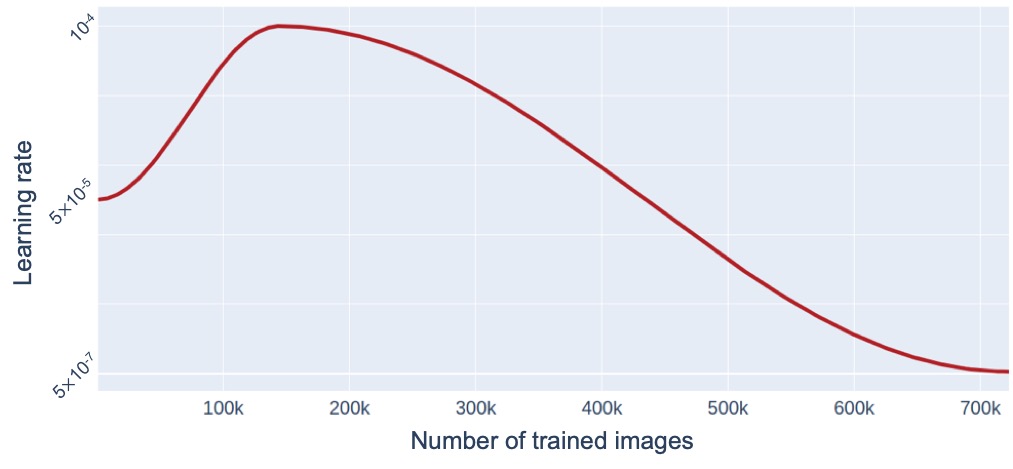}
\caption{One-cycle policy of the learning rate for training the PyNET-CA.}
\label{fig:onecycle_lr}
\end{figure}

The PyNET requires a long training time, especially at the high-resolution levels.
To address this issue and hasten convergence, we employ the one-cycle policy of the learning rate for training the PyNET-CA \cite{smith2019super}.
At each levels, the learning rate starts with $5.0 \times 10^{-5}$, reaches the maximum learning rate $1.0 \times 10^{-4}$ at the early 20\% of the training, and gradually decays to $5.0 \times 10^{-7}$ by the end of the training as in Figure \ref{fig:onecycle_lr}.
We train the model for 16 epochs per level, which is around 68\% decrement of the training epochs at the last level compared to \cite{ignatov2020replacing}.

\section{Experiment}

\subsection{Comparative studies}

\begin{figure}
\centering
\includegraphics[width=1.0\textwidth]{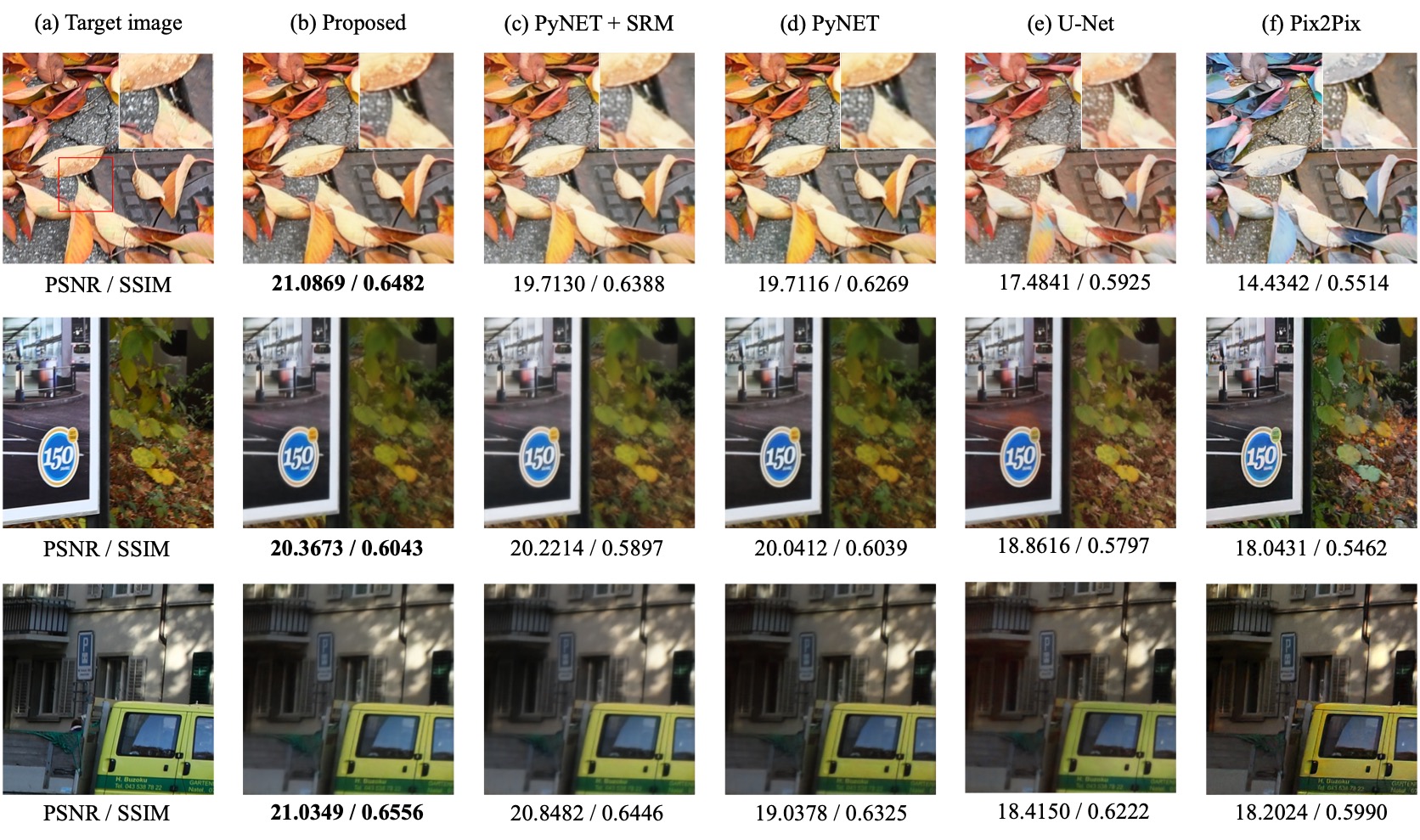}
\caption{Reconstruction result from the comparative experiments. (a) The target RGB image. (b) Reconstructed RGB image with the proposed PyNET-CA. (c)-(f) Reconstructed images from ablation studies and benchmark models.}
\label{fig:comparative}
\end{figure}

\setlength{\tabcolsep}{4pt}
\begin{table}
\begin{center}
\caption{Comparative study of the proposed method.}
\label{table:comparative}
\begin{tabular}{r c c ccc}
\hline
\noalign{\smallskip}
Model & CA & SRM & PSNR & SSIM \\
\noalign{\smallskip}
\hline
\noalign{\smallskip}
PyNET-CA (proposed)   & \checkmark & \checkmark & \textbf{21.5022} & \textbf{0.7438} \\
PyNET + SRM           &            & \checkmark & 21.4126 & 0.7375 \\
PyNET                 &            &            & 21.2071 & 0.7367 \\
\hline
U-Net                 &            &            & 20.5057 & 0.7297 \\
Pix2Pix               &            &            & 20.4502 & 0.7196  \\
\hline
\end{tabular}
\end{center}
\end{table}
\setlength{\tabcolsep}{1.4pt}

We demonstrate the effectiveness of the proposed method by comparative studies (Table \ref{table:comparative}, Figure \ref{fig:comparative}).
First, we performed ablation studies to evaluate the performance with and without the modules of the PyNET-CA.
The performance of the PyNET-CA degrades if the channel attention at the MCCA module is removed, and further degrades without the SRM \eqref{eq:srm}.
In the case of the PyNET-CA without both modules, the model architecture corresponds to the architecture of the original PyNET, network training scheme being the only difference from \cite{ignatov2020replacing}.
From the ablation studies, it can be shown that the PyNET model is enhanced in its performance with the channel attention module and the SRM upsampling.

Second, we compare the PyNET-CA model with two benchmark models, the U-Net \cite{ronneberger2015u}, and the Pix2Pix \cite{zhu2017unpaired}.
The U-Net is a encoder-decoder CNN with skip connections, applied to many image reconstruction tasks.
Although the local features of the output images were relatively well recovered, the global features such as luminance or color balance were easily lost.
The Pix2Pix is a conditional GAN for unpaired image-to-image translation tasks.
The RGB image is reconstructed with the generator, taking RAW image as the condition.
The output images were perceptually more realistic than fully supervised methods, retaining sharp edges with high-frequency details.
However, the model could not correctly reconstruct the color of the RGB images in many cases.

\subsection{AIM 2020 learned smartphone ISP challenge}

\setlength{\tabcolsep}{4pt}
\begin{table}
\begin{center}
\caption{Result of the AIM 2020 learned smartphone ISP challenge}
\label{table:challenge}
\begin{tabular}{r cc ccc}
\hline\noalign{\smallskip}
 & \multicolumn{2}{c}{Fidelity} & \multicolumn{3}{c}{Perceptual} \\
\noalign{\smallskip}
\hline
\noalign{\smallskip}
Team & PSNR & SSIM & PSNR & SSIM & MOS \\
\noalign{\smallskip}
\hline
\noalign{\smallskip}
anonymized & \textbf{22.257} & \textbf{0.7913} & 21.011 & 0.7729 & 4.0 \\
\textbf{skyb} (ours) & \textbf{21.926} & \textbf{0.7865} & 21.734 & \textbf{0.7891} & 3.8 \\
anonymized & 21.915 & 0.7842 & 21.574 & 0.7770 & \textbf{4.7} \\
anonymized & 21.909 & 0.7829 & \textbf{21.909} & 0.7829 & 4.0 \\
anonymized & 21.861 & 0.7807 & \textbf{21.861} & 0.7807 & \textbf{4.5} \\
anonymized & 21.569 & 0.7846 & 21.569 & \textbf{0.7846} & 3.5 \\
anonymized & 21.403 & 0.7834 & 21.403 & 0.7834 & 4.2 \\
anonymized & 21.179 & 0.7794 & 21.179 & 0.7794 & 4.1 \\
anonymized & 21.144 & 0.7729 & 21.144 & 0.7729 & 3.2 \\
anonymized & 20.192 & 0.7622 & - & - & - \\
anonymized & 20.138 & 0.7438 & 20.138 & 0.7438 & 2.2 \\
\hline
\end{tabular}
\end{center}
\end{table}
\setlength{\tabcolsep}{1.4pt}

We report the results from the AIM 2020 learned smartphone ISP challenge \cite{ignatov2020aim_ISP}.
There were two separate tracks which we both participated in.
The goal of the first track was to achieve highest fidelity in terms of the PSNR and the SSIM, while the goal of the second track was to reconstruct RGB images with the best perceptual quality.
In the second track, the perceptual quality was evaluated by the mean opinion score (MOS).
To find the best model with low generalisation error, we used the provided 1,204 images for validating the model during training.
We early stopped the training with highest PSNR value on the validation dataset to participate in the fidelity track.
For the perceptual track, we trained the last level for 32 epochs, and early stopped the training with lowest learned perceptual image patch similarity (LPIPS) \cite{zhang2018unreasonable} value on the validation dataset.
All models were implemented with PyTorch 1.5.0 and were trained with four NVIDIA V100 GPUs with 16GB memory each.
We applied 8x self-ensemble of 90 degree rotation and horizontal/vertical flip during test time.
We have ranked 2nd place for both the PSNR and the SSIM on the fidelity track, and 1st place for the SSIM on the perceptual track as reported in Table \ref{table:challenge}.
The challenge results demonstrate the exceptional performance of the proposed method.

\section{Conclusion}

We propose PyNET-CA, an end-to-end deep learning model for RAW to RGB image reconstruction.
The model enhances the PyNET structure to improve its performance and reduce training time.
Comparative experiments and results from the AIM 2020 challenge demonstrate the exceptional performance of the proposed model.

\section*{Acknowledgement}
This work was supported by Institute of Information \& Communications Technology Planning \& Evaluation (IITP) grant funded by the Korea government(MSIT) [2016-0-00562(R0124-16-0002), Emotional Intelligence Technology to Infer Human Emotion and Carry on Dialogue Accordingly]

\clearpage
%
%
\bibliographystyle{splncs04}

\end{document}